# Highly efficient coherent amplification of zero-field spin waves in YIG nano-waveguides

K. O. Nikolaev[1], S. R. Lake[2], B. Das Mohapatra[2], G. Schmidt[2,3], S. O. Demokritov[1], and V. E. Demidov[1*]

[1]*Institute of Applied Physics, University of Muenster, 48149 Muenster, Germany*

[2]*Institut für Physik, Martin-Luther-Universität Halle-Wittenberg, 06120 Halle, Germany*

[3]*Interdisziplinäres Zentrum für Materialwissenschaften, Martin-Luther-Universität Halle-Wittenberg, 06120 Halle, Germany*

Transmission and processing of information at the nanoscale using spin waves and their quanta – magnons, offers numerous advantages and opportunities that make it a promising next-generation technology for integrated electronics. The main challenges that still need to be addressed to ensure high competitiveness of magnonic devices include finding ways to efficiently amplify spin waves in nanostructures and developing nanocircuits that can operate without the need for an external bias magnetic field. Here we demonstrate how these two challenges can be solved using nano-waveguides fabricated from a low-loss magnetic insulator. We show that by using local parametric pumping with a power of only a few milliwatts, one can achieve coherent amplification of spin-wave pulses by more than two orders of magnitude at zero bias magnetic field. Our results provide a simple solution to problems that have long prevented the implementation of efficient integrated magnonic circuits.

The emergent field of magnonics considers the use of spin waves propagating in magnetic nanostructures and networks for low-energy transmission of signals and processing of information[1]. Owing to short wavelengths, down to a few tens of nanometers[2-4], and excellent controllability by magnetic and electric fields[5-10], spin waves offer many important opportunities for implementation of efficient integrated devices and circuits[11-13]. Moreover, due to the rich spectrum of nonlinear phenomena[14-18], spin waves provide a unique platform for advanced information processing including reservoir and neuromorphic computing[19-21].

The benefits of spin waves are significantly compromised by two important factors. First, in nanostructured magnonic waveguides, spin waves exhibit relatively short propagation lengths. Even in nano-structures fabricated from ultra-low-loss magnetic insulator – Yttrium Iron Garnet (YIG)[22-24], the propagation length typically does not exceed several tens of micrometers[25-27]. Therefore, the implementation of complex magnonic nano-circuits constituted by many cascaded devices necessarily requires a periodic recovery/amplification of spin-wave signals. Although this problem is well recognized, until now there was no efficient solution for amplification of spin waves at the nanoscale. Recently it was shown that the nanoscale amplification can be achieved using spin-transfer torque phenomena[26], stimulated magnon scattering[25], or bistability phenomena[27]. However, the achieved gains did not exceed a factor of 5-6. Additionally, the last two mechanisms lack coherence, i.e. the important phase information is lost during amplification.

Another mechanism, which can be used for amplification of spin waves is the parametric pumping[28]. This mechanism was intensively studied in the past in macroscopic millimeter-scale spin-wave devices and was shown to enable large gains up to a factor of several hundreds[29-32]. However, previous attempts to apply this mechanism to microscopic structures have had limited success. Reasonable parametric gains could only be observed in transition-metal ferromagnetic structures[33], where the achieved amplification was completely compromised by large losses typical for these materials. In contrast, in nanostructures fabricated from low-loss magnetic insulators, such as YIG, the parametric mechanism could only be used to generate spin waves[34], while no parametric amplification has been demonstrated so far.

In addition to the missing efficient amplification of spin waves, nano-magnonics also strongly suffers from the requirement to apply static bias magnetic fields, which is needed to operate most of the spin-wave devices. This requirement significantly complicates the practical implementation of devices and circuits, as it requires the integration of space- and energy-consuming magnetic systems. Accordingly, considerable efforts have been directed towards finding ways to achieve propagation of spin waves at zero magnetic field. Although a number of approaches to solving this problem have been demonstrated[35-38], most of them require the use of magnetic materials with large magnetic losses. Only recently it was shown[39] that zero-field transmission of spin waves can be realized in nanosized waveguides fabricated from low-loss YIG films. Such waveguides have been proven to exhibit a highly stable uniform magnetic state at zero bias magnetic field and support the propagation of spin waves with long propagation lengths. These features make such waveguides promising building blocks for the development of complex magnonic devices and networks.

In this work, we experimentally demonstrate that spin waves propagating in YIG nano-waveguides at zero bias magnetic field can be efficiently amplified using a parametric pumping localized in an active region as small as 2 µm in length. We show that a pumping power of only 4 mW is sufficient to fully compensate for the losses experienced by spin-wave pulses over the propagation distance of 10 µm. At the highest power of 15 mW used in our experiment, we achieve gains of more than two orders of magnitude, which are mainly limited by the onset of nonlinear phenomena at large intensities of the amplified pulses. The achieved gains are sufficient to completely compensate the spin-wave losses at propagation distances of more than 60 µm using one single local amplifier. Using phase-resolved measurements, we directly prove that the studied amplification process does not disturb the phase coherence of spin-wave pulses. Therefore, it can also be used in advanced magnonic circuits taking advantage of the possibility of encoding the information into the phase of spin waves and using interference phenomena for information processing.

**RESULTS**

**Studied system and approach.**

Figure 1 schematically illustrates the idea of the experiment. Spin-wave pulses with the carrier frequency $f_s$ propagate in a 500-nm wide and 80-nm thick waveguide fabricated from a YIG film prepared by the pulsed laser deposition[24]. The measurements are performed at zero static magnetic field. Under these conditions, the static magnetization in the waveguide is spatially uniform and is aligned parallel to the axis of the waveguide[39], which corresponds to the propagation configuration of the so-called backward-volume spin waves[28]. Spin waves are inductively excited by applying 20-ns long pulses of microwave current through a 500-nm wide and 200-nm thick input Au antenna. After traveling 15 μm along the waveguide, the pulses encounter another antenna with a width of 2 μm. This antenna is used for parametric amplification. To implement the amplification, we apply to this antenna microwave pulses at a frequency that is twice the frequency of the signal carried by spin waves – the parametric pumping. Dynamic magnetic field at this frequency cannot linearly excite magnetic dynamics. Instead, it serves as a periodic modulation of the system parameters, which is known to lead to compensation of magnetic damping[29-31]. At sufficiently high powers of the pumping signal, the damping becomes negative, which leads to an amplification of the spin-wave pulse passing through the pumping region.

We study the propagation and amplification of the spin-wave pulses using space- and time-resolved micro-focus Brillouin light scattering (BLS) spectroscopy[40]. We focus probing laser light into a diffraction-limited spot on the YIG waveguide (see Fig. 1) and analyze the modulation of this light due to its scattering from spin waves. The detected BLS intensity is proportional to the spin-wave intensity at the position of the probing-light spot, which provides the opportunity to directly observe the propagation of spin-wave pulses with high spatial resolution.

**Evidence of parametric amplification.**

Figure 2 illustrates the main achievement of our work – highly efficient local amplification of spin-wave pulses. Here we use signal pulses with the frequency $f_s$=1.5 GHz, which corresponds to the middle of the spin-wave band at zero static magnetic field (see Supplementary information). First, we analyze the propagation of pulses without applying parametric pumping (Figs. 2a and 2b). We move the probing light spot along the waveguide with a step of 500 nm and record the temporal traces of the BLS intensity at different $z$-positions. The results of these measurements are presented in Fig. 2a as a color map in space-time coordinates. Representative temporal profiles of the spin-wave pulses recorded at points A ($z = 0$), B ($z = 15.5$ μm), and C ($z = 30$ μm) are shown in Fig. 2b. As seen from these data, despite the ultimately low magnetic damping in YIG, the spin-wave pulse experiences a noticeable decay as it propagates in the waveguide: after a travel distance of 30 μm, its intensity decreases by about a factor of 10. From the slope of the dashed line in Fig. 2a we determine the group velocity of the spin waves – $v_g = 0.36$ μm/ns. At such velocity, the center of the pulse passes through the 2-μm wide pumping region in 6 ns. Accordingly, we chose the duration of the pumping pulse to be 26 ns, so that the pumping starts when the leading edge of the 20-ns wide signal pulse reaches the pumping region, and ends when its trailing edge leaves this region.

Figures 2c and 2d show the results obtained when applying a signal with a frequency $f_p = 2f_s = 3.0$ GHz and a power of $P_p = 10$ mW to the pumping antenna. Simple comparison of these data with the data in Figs. 2a and 2b clearly indicates the high efficiency of the parametric amplification. In particular, the intensity of the pulse at point B (directly behind the pumping antenna) increases by a factor of about 16. Moreover, the intensity of the pulse at point C ($z = 30$ μm) is now larger than the intensity of the initially excited pulse at point A ($z = 0$). In other words, in this representative example, the complete compensation of spin-wave losses corresponding to the propagation distance of 30 μm is achieved.

The data of Fig. 2c also show the excitation by the parametric pumping of the so-called idle pulse, which is typical for all parametric amplifiers. The parametric amplification can be viewed as the stimulated splitting of the photon of the pumping field into two magnons with half the frequency[28]. In order to satisfy the conservation of momentum, these two magnons must have opposite wave vectors, i.e., propagate in opposite directions. This leads to a formation of the idle pulse propagating oppositely to the amplified signal pulse. Note that, although the idle pulse is a byproduct of the parametric amplification process, it can also be used for certain applications[41].

To better quantify the demonstrated amplification, we show in Fig. 3 the spatial dependences of the peak intensity of the initial and the amplified pulses. As seen from these data, the intensity always shows a clear exponential decay (note the logarithmic scale of the vertical axis) characterized by the propagation length of 28 μm, which is typical for zero-field backward volume waves in the waveguides used[39]. Extrapolating the exponential fit of the data for the initial pulse beyond the pumping region (dashed line in Fig. 3), one can see that the presence of the pumping antenna results in a small drop in intensity. This drop is due to the unavoidable reflection and absorption of spin-wave energy by the pumping antenna. We emphasize that this drop is insignificant compared to the increase in intensity achieved using the parametric pumping.

**Power dependences of the parametric gain**

We now discuss how the observed amplification depends on the pumping and signal power. Figure 4a shows the dependences of the BLS signal detected at the output of the amplifier (point B in Fig. 2a) as a function of the pumping power $P_p$ for two different powers of the signal pulse $P_s$, as labelled. The data in Fig. 4a show that the intensity of the amplified pulse increases approximately exponentially (note the logarithmic scale of the vertical axis) with increasing pumping power $P_p$ < 10 mW and then tends to saturate. Note that at the maximum pumping power used in the experiment ($P_p$ = 15 mW) the intensity of the amplified pulses becomes almost independent of the power of the signal pulse. This indicates that the maximum intensity of the amplified pulse is limited by the onset of nonlinear phenomena, such as the nonlinear damping and nonlinear shift of the spin-wave spectrum leading to the reduction of the efficiency of the interaction of spin waves with the pumping field[28]. This nonlinear limitation results in a dependence of the parametric gain on the intensity of the initial pulse at large $P_p$. In particular, at $P_p$ = 15 mW, the gain is limited to about 28, if the input power of $P_s$ = 0.1 mW is used. However, when the input power is reduced to $P_s$=0.05 mW, an almost twice larger gain (≈55) can be achieved.

Figure 4b characterizes the dependence of the parametric gain on the input power in more detail. At small $P_p$ = 7 mW far below saturation, the gain is almost independent of the signal power. At $P_p$ = 10 mW, the gain already decreases noticeably as the signal power increases from 0.01 to 0.2 mW. Finally, at $P_p$ = 15 mW, the gain becomes very sensitive to the signal power. Note that at the smallest $P_s$=0.01 mW, the gain reaches a value of more than 100. These observations allow us to formulate guidelines for the optimum use of parametric amplification in magnonic circuits. In particular, they show that the efficiency of parametric amplification is particularly high in the small-signal regime. In practice, such a situation is realized, for example, if local amplification is used to restore strongly attenuated spin-wave pulses that have travelled a long distance. Given the measured propagation length of 28 μm, the intensity of the initially excited spin-wave pulse decreases by a factor of 100 after traveling 64 μm. It can be expected that under these conditions the signal intensity is small enough to achieve gains greater than 100 using $P_p$ = 15 mW. In other words, by placing parametric amplifiers every 64 μm and using the pumping with the power of 15 mW, one can completely compensate the propagation losses of spin waves in a spatially extended magnonic circuit. Depending on requirements of particular application, one can also place amplifiers with smaller spacing and use lower pumping powers. For example, if the amplifiers are placed every 10 μm, the propagation losses can be completely compensated using the power as small as $P_\text{p}$ = 4 mW.

**Coherence of the amplification process.**

One of the important advantages of wave-based transmission and processing of information is that the information can be encoded not only into intensity, but also into the phase of the waves. Therefore, a key requirement for the amplification/recovery process is that it does not lead to loss of phase information. To show that this is the case for the parametric process under study, we perform BLS measurements with phase resolution[40]. To achieve sensitivity of the BLS apparatus to the phase of spin waves, we exploit the interference of the light scattered from the spin waves with reference light modulated by the microwave signal used to excite the input spin-wave pulses. The resulting signal is then proportional to $A_\text{sw}(z)\cos(\Delta\varphi(z))$, where $A_\text{sw}$ is the amplitude of the spin wave at the observation point $z$, and $\Delta\varphi$ is the shift of its phase relative to the phase of the microwave signal at the input antenna.

Figure 5a shows the phase-resolved spatial profiles of the spin-wave pulses recorded at times $t$ = 30 and 85 ns. At $t$ = 30 ns, the spin-wave pulse is located between the input antenna and the pumping antenna, i.e., it has not yet undergone amplification. At $t$ = 85 ns the entire amplified pulse has already left the pumping region. As seen from the data in Fig. 5a, both the initial and amplified pulses exhibit clear periodic oscillations, which are determined by the oscillations of the term $\cos(\Delta\varphi(z))$. This fact already indicates that the phase of the amplified pulse is locked to that of the initial pulse. Indeed, since in BLS measurements the signal is averaged over a very large number of pulses, the term $\cos(\Delta\varphi(z))$ would average to zero if there were no phase coherence. Note that such averaging is clearly seen for the idle pulse: although the intensities of the amplified and the idle pulse are almost the same (see Fig. 2d), we do not observe in Fig. 5a an oscillating signal on the left from the pumping antenna indicating that the phase of the idle pulse is not locked to that of the initial pulse. Instead, we observe a weak non-oscillating signal. The latter is caused by a small contribution to the measured signal of the signal proportional to the spin-wave intensity, which is difficult to completely exclude. We emphasize that this result is consistent with the concept of the parametric amplification process as a stimulated splitting of a pumping-field photon into two magnons. It is known that such a process conserves the sum of the phases of the three involved quasiparticles[28]. Since the splitting is stimulated by the signal pulse, the phase of the generated magnon contributing to the amplified pulse is automatically synchronized with the phase of the signal. However, since the pumping field is not phase-locked to the input signal, the phase of the idle magnon is undefined.

The phase-resolved profiles in Fig. 5a also allow us to directly determine the wave number $k_0$ of the carrier wave for the initial and amplified pulses. We use Fourier transformation of the data and obtain for both pulses $k_0 = 3.98$ μm$^{-1}$, which corresponds to the wavelength of 1.58 μm. This result indicates that the amplification process not only preserves the phase of the signal, but also does not affect the wavelength of the carrier spin wave.

We now use phase-resolved measurements to demonstrate that the observed intensity gain (Fig. 4b) is entirely due to the amplification of the coherent component of the signal pulse. To do this, we sweep the phase of the microwave signal applied to the input antenna $\varphi_0$ from 0 to 360 degree keeping the phase of the reference modulation constant, and carry out interference measurements at point B ($z = 15.5$ μm) at the output of the parametric amplifier. In these measurements we use the lowest signal power ($P_s = 0.01$ mW) and the highest pumping power ($P_p = 15$ mW), which corresponds to the case of the ultimate gain of 147 observed in the intensity measurements (Fig. 4b). Figure 5b shows the resulting interference curves recorded with and without pumping. As expected, both of them exhibit sinusoidal oscillation. The amplitude of this oscillation is proportional to the amplitude of the coherent component, phase-locked to the input microwave signal. We fit the data with a sinusoidal function and find the ratio between the oscillation amplitudes of 11. This corresponds to the intensity gain of 121, which agrees well with the value observed in the intensity measurements. This clearly indicates that the obtained gains indeed characterize the amplification of the coherent component, and that the amplification process remains highly coherent even for the weakest signals used in our experiments.

In conclusion, we have experimentally demonstrated a highly efficient approach to address the key challenges in nano-magnonics. We show that spin waves propagating at zero bias magnetic field in waveguides made from low-loss magnetic insulators can be efficiently amplified by local parametric pumping and that this amplification process preserves the phase information carried by spin waves. Our results open new avenues for the realization of complex magnonic circuits for nano-scale transmission and processing of information, which can operate without a bias magnetic field and do not require energy-inefficient conversion of spin waves into electric currents to recover spin-wave signals.

**Methods**

**Sample fabrication.** The YIG waveguides are fabricated using electron beam lithography, pulsed laser deposition, and lift-off. Electron beam lithography is performed using a two-layer PMMA resist spin-coated on a <111> oriented GGG substrate. The PMMA is covered by an ultra-thin gold film to provide a conducting surface and to avoid charging during the exposure. After electron beam exposure (30 kV) the gold is removed using a potassium iodide solution. The PMMA is developed in pure isopropanol. Possible residual resist in the developed areas is removed using oxygen plasma. A nominally 100 nm thick YIG film is deposited at room temperature by pulsed laser deposition using a process developed by Hauser and co-workers[24]. After lift-off in acetone the sample is annealed in oxygen for 3 hours at 800 degrees. Wet etching in phosphoric acid removes approx. 20 nm of YIG and provides smooth edges. For high frequency measurements, microstrip antennas are deposited on top of the YIG waveguides using a similar process, but replacing the PLD of YIG by e-gun evaporation of 10 nm of titanium and 200 nm of gold, and discarding the annealing step after lift-off.

**Micro-focus BLS measurements.** All measurements are performed in a pulsed regime. The width of the pulses used for the excitation of spin waves is 20 ns, and their repetition period is 500 ns. To detect propagating spin-wave pulses with spatial and temporal resolution, we use inelastic scattering of laser light from spin waves. Light with a wavelength of 473 nm and a power of 0.25 mW is generated by a single-frequency laser with a linewidth < 5 MHz. We use a high-performance microscope objective lens with a magnification of 100 and a numerical aperture of 0.9 to focus the laser light into a diffraction-limited spot, which ensures high spatial resolution of the measurements. The position of the probing spot is monitored and actively stabilized using a home-built digital microscope. The scattered light is analysed using a six-pass Fabry-Perot interferometer and a high-sensitivity single-photon detector. By determining the delay of the detected photons of the scattered light relative to microwave pulses used to excite spin waves, we achieve a temporal resolution of about 1 ns. To obtain sensitivity of the technique to the phase of spin waves, we use the interference of scattered light with the reference light. The reference light is modulated at the frequency of spin waves using a broadband electro-optical modulator.

**ACKNOWLEDGEMENTS**

This work was supported by the Deutsche Forschungsgemeinschaft (DFG, German Research Foundation) – project number 529812702 (K.O.N., G.S., V.E.D). The work of S.O.D. was supported by the Deutsche Forschungsgemeinschaft (DFG, German Research Foundation) – SFB 1459/2 2025 – 433682494.

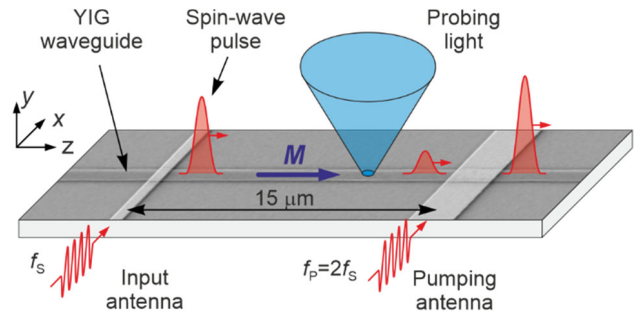

**Figure 1 | Schematics of the experiment.** Spin-wave pulses with the carrier frequency $f_s$ propagate in a 500-nm wide and 80-nm thick YIG waveguide at zero bias magnetic field. The pulses are amplified using pulsed parametric pumping field with a frequency $f_p = 2f_s$, which is generated by a 2-µm wide pumping antenna. The propagation and amplification of the spin-wave pulses is studied with spatial and temporal resolution using Brillouin light scattering spectroscopy.

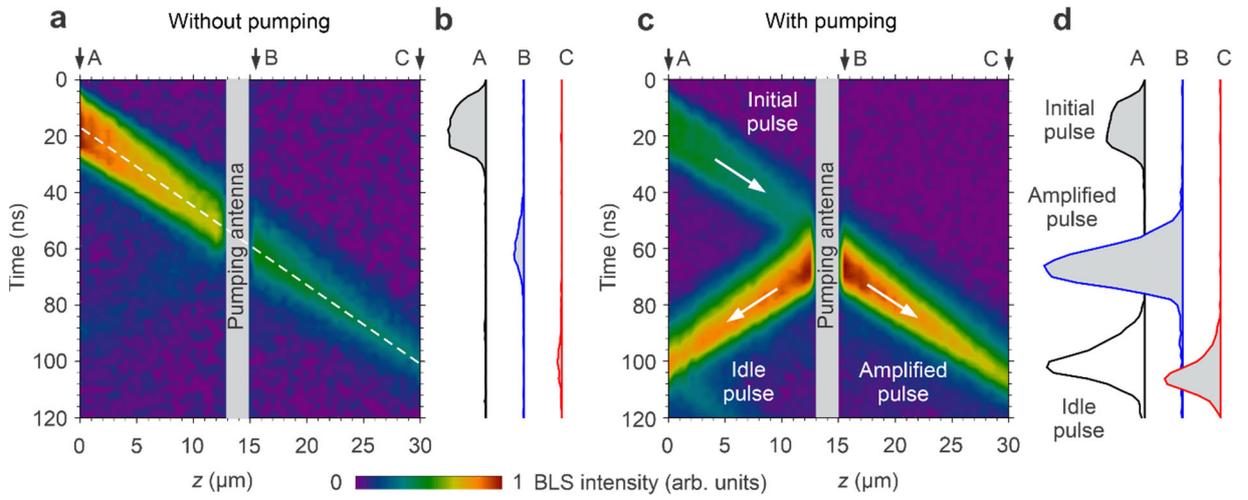

**Figure 2 | Propagation and amplification of spin-wave pulses.** (a) and (c) Color maps of spin-wave intensity in space-time coordinates. (a) shows the data obtained without pumping. (c) shows the data obtained with pumping with a power of $P_p = 10$ mW. (b) and (d) show representative temporal profiles of spin-wave pulses recorded at points A ($z = 0$), B ($z = 15.5$ μm), and C ($z = 30$ μm) without (b) and with (d) pumping. Dashed line in (a) marks the propagation of the center of the spin-wave pulse. The slope of the line corresponds to the group velocity of $v_g = 0.36$ μm/ns. The data are obtained at zero bias magnetic field. Frequency of the spin-wave pulses is $f_s = 1.5$ GHz. Power of the signal applied to the input antenna is $P_s = 0.1$ mW.

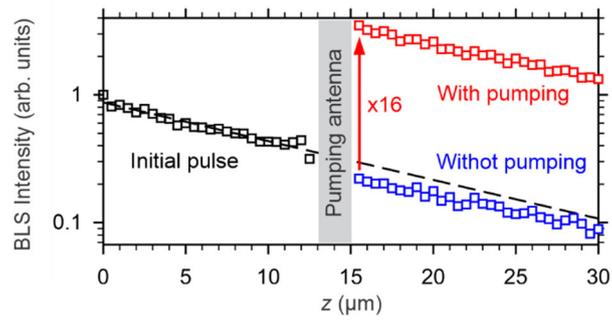

**Figure 3 | Quantitative characterization of the amplification.** Spatial dependences of the peak intensity of spin-wave pulses with and without pumping. The dashed line shows the exponential fit of the data in the range $z = 0 - 12$ μm. The slope of the line corresponds to the propagation length of spin waves of 28 μm. The arrow illustrates the increase in intensity by a factor of 16 due to parametric amplification. The data are obtained at zero bias magnetic field. Frequency of the spin-wave pulses is $f_s = 1.5$ GHz. Power of the signal applied to the input antenna is $P_s = 0.1$ mW. Pumping power is $P_p = 10$ mW.

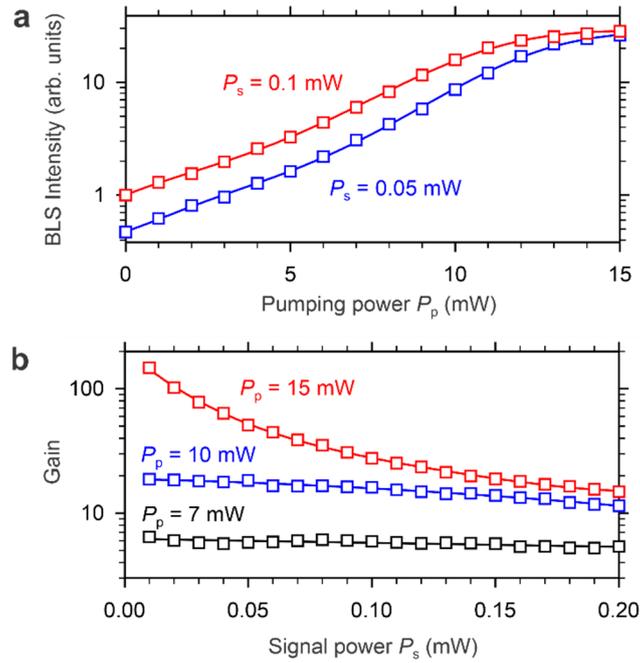

**Figure 4 | Power dependences of the parametric amplification.** (a) Dependences of the BLS signal detected at the output of the amplifier (point B in Fig. 2a) as a function of the pumping power $P_p$ for different powers of the signal pulse $P_s$, as labelled. (b) Dependences of the parametric gain on the signal power $P_s$ for different pumping powers $P_p$, as labelled. Symbols show the experimental data. Curves are guides for the eye. The data are obtained at zero bias magnetic field. Frequency of the spin-wave pulses is $f_s = 1.5$ GHz.

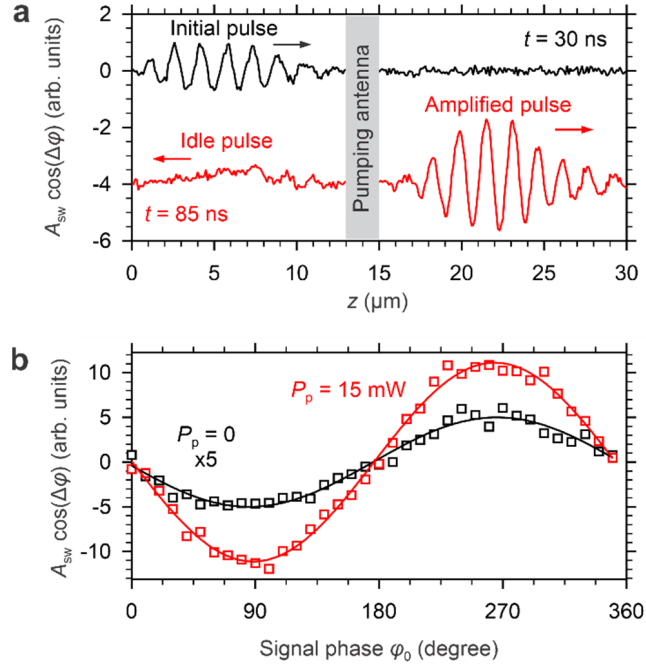

**Figure 5 | Evidence for the coherence of the amplification process.**
(a) Phase-resolved spatial profiles of the spin wave pulse recorded before ($t$ = 30 ns) and after ($t$ = 85 ns) amplification. Curves are vertically shifted for clarity. The data are recorded at $P_s$ = 0.1 mW and $P_p$ = 10 mW. (b) Dependences of the phase-resolved signal on the phase of the microwave signal applied to the input antenna, recorded at the output of the amplifier with and without pumping. The data are recorded at the lowest used signal power $P_s$ = 0.01 mW and the highest used pumping power $P_p$ = 15 mW. Symbols show the experimental data. Curves are the fit of the data by a sinusoidal function. The data are obtained at zero bias magnetic field. Frequency of the spin-wave pulses is $f_s$ = 1.5 GHz.